\documentclass[aps,eqsecnum,prd,twocolumn]{revtex4}
\usepackage{graphics,graphicx}
\usepackage{amsmath}
\usepackage{amssymb,latexsym,mathrsfs}
\usepackage{hyperref}

\def\bea{\begin{eqnarray}}
\def\eea{\end{eqnarray}}
\def\ba{\begin{array}}
\def\ea{\end{array}}

\def\beq{\begin{equation}}
\def\eeq{\end{equation}}

\begin{document}

\title{Temperature dependent weak value of dwell time for a two state particle tunneling through a thermal magnetic barrier}

\author{Samyadeb Bhattacharya \footnote{sbh.phys@gmail.com}}
\affiliation{Physics and Applied Mathematics Unit, Indian Statistical Institute, 203 B.T. Road, Kolkata 700 108, India \\}

\vspace{2cm}
\begin{abstract}

\vspace{1cm}

Dwell time for a two state particle tunneling through a noisy thermal magnetic barrier has been calculated by studying the time evolution of the system. The effect of temperature has been included by averaging over the environmental magnetic modes. The time scale has been calculated in the framework of weak measurement. The dwell time initially increases with the rise of temperature and finally saturates. The increment of dwell time can be explained by the phenomena of quantum memory loss caused by efficient energy exchange with the environmental modes. The saturation region at higher temperature corresponds to the process of thermal hopping of the barrier.
\vspace{2cm}


\end{abstract}

\vspace{1cm}

\maketitle

\section{Introduction}
The paradigm of quantum tunneling explains diverse processes from nuclear alpha decay to current flow in Josepshon junctions. The phenomena of a quantum particle tunneling through a barrier with such a potential height that is classically unsurmountable, is actually quite well understood through the work of Gamow and many others in the later period. But after so many years of intense theoretical and experimental study, we still do not fully understand the aspect of ``tunneling time"; ie how much time does a particle take to cross a classically unsurmountable barrier. This problem has quite a long history \cite{1}. It is related to the fundamental question of introducing time as a quantum mechanical observable. More than five decades ago, the problem was theoretically addressed by Hartman \cite{2}. Hartman analyzed the temporal aspect of tunneling to infer certain very important properties of the transmitted wave packet. He found that for the case of opaque barrier, the tunneling time is independent of barrier thickness and it seems to violate the relativity postulate. To explain this ambiguity, some suggestions have been made \cite{3,4,5,6} for the intuitive understanding of the physical aspect of tunneling time, especially group delay and dwell time \cite{1}. In those works, it has been stated that the group delay, which is also directly related to the Dwell time by an additive self-interference term, is not the time taken by the wave packet to travel through the barrier region, but practically the lifetime of stored energy or particle leaking through the barrier at both ends. For such situations where the surroundings of the barrier is dispersionless, the dwell time equals to the phase time because of the fact that the self-interference term vanishes in those cases. In those situations, the dwell time represents the lifetime of energy storage in the barrier region. In a recent work \cite{7}, the present authors have considered the case of quantum tunneling, where the system in concern is coupled to a dissipative environment and found that, in presence of dissipation, the dwell time depends on the barrier thickness to show  quasi-classical behavior. What we have concluded there is that the continuous interaction with the environmental bath modes makes the behavior quasi-classical. But in that work, we have dealt only with the dynamics of the transmitting particle through the approach of Quantum-Langevin equation and did not get into the nature of the coupling with the environmental modes. In the Quantum Lengevin equation, the environmental coupling is incorporated by the inclusion of the dissipation coefficient and the random force, which are related to the parameters of the environmental modes. To study the true nature of dissipation, we have to investigate the coupling parameters in a more rigorous manner. In the present work, we have considered the particular situation, where a spin half particle is tunneling through a thermal magnetic barrier. We have studied the temporal evolution of the quantum state of the particle in an equilibrium situation to find that the decay rate depends on the temperature along with other parameters of the environmental modes. At high temperature, thermal activation dominates and the particle hops over the intervening barrier. At low temperature, the process of tunneling becomes more and more important. This feature can be inferred from the result of our calculations. \\
Here we will focus on the approach of weak measurement to derive the appropriate expression for the dwell time. Aharonov et.al \cite{8} along with other authors \cite{9} have previously dealt with the idea of tunneling time (especially the dwell time) from this particular context. In fact the idea of the weak measurement of an observable in quantum mechanical framework, was originally introduced to the community by the works of Aharanov et.al \cite{9a,9b,9c}. This quantity is the result of a standard measurement, performed upon a pre and post selected (PPS) ensemble of quantum systems, leaving the interaction between the measurement apparatus and each system sufficiently weak. Unlike the usual case of strong measurement, this specific type of measurement of an observable for a PPS system does not appreciably disturb the quantum system. Here we interpret ``measurement" as the interaction of the environmental thermal magnetic field with the system in concern. This interaction is of course weak in our consideration. In a recent work \cite{10}, we have also considered the approach of weak measurement in the problem of tunneling time in a dissipative environment. We will mainly follow the same framework. But here we will treat the environmental interaction in much detail to get to a specific expression of decay constant which, as we will show later, has got a very important role in our final derivation. In the next section, we will present the theoretical framework of the two level system interacting with a thermal magnetic field and derive the expression of decay constant. Then in the 3rd section, we will derive the dwell time in the framework of weak measurement and use the expression of decay constant to arrive at our final result. After that we will conclude with some possible implications.

\section{Expression of decay constant for two level system interacting with a thermal field}

During the last decades or so, after the aberration-corrected transmission electron microscopy (TEM) \cite{11,12} came into picture, resolving power of electron microscopy has been improved considerably. But we are still faced with the problem of coherence loss in the process of image formation. In a very recent work \cite{13}, the cause of this decoherence has been shown to be the influence of the magnetic field noise caused by the thermally driven currents in the conducting material of the focussing elements. We are considering a situation, where the two level particle (electron for example) is undergone the influence of such a thermal magnetic field. \\
The system can be described by the Hamiltonian
\beq\label{1}
H_s=\frac{1}{2}\hbar \Omega \sigma_z
\eeq

where $\sigma_z$ is the Pauli spin matrix in usual notation. The interaction Hamiltonian can be introduced as \cite{14}
\beq\label{2}
H_i=g(\sigma_{+}+\sigma_{-})B
\eeq

where $B$ is the external thermal magnetic field and $g$ is the coupling parameter. The total Hamiltonian for the system and the coupled magnetic field can be expressed as

\beq\label{3}
H_T=H_s+H_f+H_i
\eeq
$H_f$ is the reservoir Hamilton, which can be represented by collection of harmonic oscillators.

\beq\label{3a}
H_f=\sum_n \hbar\omega_n \hat{a}_n^{\dag}\hat{a}_n
\eeq

where $\hat{a}_n$ and it's hermitian conjugate $\hat{a}_n^{\dag}$ are the annihilation and creation operator respectively. In presence of dissipation and noise, the dynamics can be represented by the equation

\beq\label{3b}
\dot{\hat{a}}_n=-i\omega_n \hat{a}_n -\eta \hat{a}_n +\sqrt{2\eta}\zeta (t)
\eeq
where $\zeta(t)$ and $\eta$ represents noise and dissipation respectively. Similarly we can find the evolution of $\hat{a}^{\dag}_n$ \cite{14}. \\
As we have stated in the introduction, the interaction with the field is assumed to be weak, so that the two level system and the field do not interact to the first approximation. To solve the von Neumann's equation, we investigate through the interaction picture where the density operator for the particle-field interacting system can be expressed as
\beq\label{4}
i\hbar\dot{\rho}_{i}(t)=[H_i(t),\rho_s(t)]
\eeq

where

\beq\label{5}
H_i(t)=\exp\left(-\frac{H_s+H_f}{i\hbar}t\right)H_i\exp\left(\frac{H_s+H_f}{i\hbar}t\right)
\eeq

As the system Hamiltonian and field Hamiltonian represents different degrees of freedom, they commute. So the time evolution of the interaction Hamiltonian will be

\beq\label{6}
H_i(t)=g\left(\sigma_{+}e^{i\Omega t}+\sigma_{-}e^{-i\Omega t}\right)B(t)
\eeq

where

\beq\label{7}
B(t)=\exp\left(-\frac{H_f}{i\hbar}t\right)B\exp\left(\frac{H_f}{i\hbar}t\right)
\eeq

This is the field operator in the interaction picture. This operator contains a wide range of frequencies, but among them only the ones are almost in resonance with $\pm \Omega$ are important. These frequencies are sufficiently slowly varying to last for significant time, while the others oscillate so rapidly their net effect can be neglected. Under this approximation, the correlation function for the fields can be written as \cite{14}

\beq\label{8}
e^{i\Omega(t-t')}\langle B(t)B(t') \rangle\sim 4\hbar \Omega (N(\Omega)+1)\delta(t-t')
\eeq

Similarly

\beq\label{9}
e^{-i\Omega(t-t')}\langle B(t)B(t') \rangle\sim 4\hbar \Omega N(\Omega)\delta(t-t')
\eeq

where $N(\Omega)$ is the Planck function, given by

\beq\label{10}
N(\Omega)=\frac{1}{\exp\left(\frac{\hbar \Omega}{KT}\right)-1}
\eeq

So Eqn.(\ref{4}) representing the von Nuemann equation for the interaction density operator $\rho_i$, is now a kind of quantum white noise equation containing much more information than we actually want from our system, since it contains all the information about the field too. So we need to obtain an equation for a reduced density operator after tracing out over the field variables. The master equation for the reduced density matrix $\widetilde{\rho_i}$ can be expressed as \cite{14}

\beq\label{11}
\begin{array}{ll}
\frac{d\widetilde{\rho_i}}{dt}=\frac{2g^2\Omega}{\hbar} (N(\Omega)+1)\left[\sigma_{-}\widetilde{\rho_i}\sigma_{+}-\sigma_{+}\sigma_{-}\widetilde{\rho_i}-\widetilde{\rho_i}\sigma_{+}\sigma_{-}\right]\\
~~~~~~ +\frac{2g^2\Omega}{\hbar}N(\Omega)\left[\sigma_{+}\widetilde{\rho_i}\sigma_{-}-\sigma_{-}\sigma_{+}\widetilde{\rho_i}-\widetilde{\rho_i}\sigma_{-}\sigma_{+}\right]

\end{array}
\eeq

Let us now consider a slightly more intricate and general situation, where we take a complex electromagnetic field of the form

\beq\label{12}
B(t)\rightarrow B(t)+\Lambda e^{i\Omega t}+\Lambda^{*} e^{-i\Omega t}
\eeq

For this case the master equation for the density operator ($\rho_i$) is modified as

\beq\label{13}
\begin{array}{ll}
\frac{d\widetilde{\rho_i}}{dt}=\frac{2g^2\Omega}{\hbar} (N(\Omega)+1)\left[\sigma_{-}\widetilde{\rho_i}\sigma_{+}-\sigma_{+}\sigma_{-}\widetilde{\rho_i}-\widetilde{\rho_i}\sigma_{+}\sigma_{-}\right]\\
~~~~~~ +\frac{2g^2\Omega}{\hbar}N(\Omega)\left[\sigma_{+}\widetilde{\rho_i}\sigma_{-}-\sigma_{-}\sigma_{+}\widetilde{\rho_i}-\widetilde{\rho_i}\sigma_{-}\sigma_{+}\right]\\
~~~~~~-\frac{ig}{\hbar}\left[\left(\Lambda \sigma_{+}+\Lambda^{*}\sigma_{-}\right),\widetilde{\rho_i}\right]

\end{array}
\eeq

Solving this master equation and after doing some algebra, we get the time evolution for the expectation values of the Pauli matrices as

\beq\label{14}
\begin{array}{ll}
\frac{d\langle\sigma_{+}\rangle}{dt}=-\frac{2g^2 \Omega}{\hbar}(2N(\Omega)+1)\langle\sigma_{+}\rangle-\frac{ig}{\hbar}\Lambda^{*}\sigma_z \\
\frac{d\langle\sigma_{-}\rangle}{dt}=-\frac{2g^2 \Omega}{\hbar}(2N(\Omega)+1)\langle\sigma_{-}\rangle+\frac{ig}{\hbar}\Lambda\sigma_z \\
\frac{d\langle\sigma_z\rangle}{dt}=-\frac{4g^2 \Omega}{\hbar}(2N(\Omega)+1)\langle\sigma_z\rangle-\frac{4g^2 \Omega}{\hbar}\\
~~~~~~~~~~-\frac{i}{2}\frac{g}{\hbar}(\Lambda\langle\sigma_{+}\rangle-\Lambda^{*}\langle\sigma_{-}\rangle)
\end{array}
\eeq

Solving for stationary situation, we can get

\beq\label{15}
\begin{array}{ll}
\langle\sigma_z\rangle=\frac{-(2N(\Omega)+1)}{(2N(\Omega)+1)^2+2|\Lambda|^2/g^2\Omega^2}\\
\langle\sigma_{+}\rangle=\langle\sigma_{-}\rangle^{*}=\frac{2i\Lambda/g\Omega}{(2N(\Omega)+1)^2+2|\Lambda|^2/g^2\Omega^2}
\end{array}
\eeq

Consider the control Hamiltonian of the system as

\beq\label{16}
H_s= \frac{1}{2}\hbar\Omega \sigma_z-\hbar\Delta\sigma_x
\eeq

$\sigma_x$ acts as a small perturbation in such a way that it does not disturb the stationary situation quite significantly. According to this Hamiltonian, the time evolution of the system is expressed by the unitary operator

\beq\label{17}
U_s(t)=\exp\left[-i\left(\frac{\Omega}{2}\sigma_z-\frac{\Delta}{2}\left(\sigma_{+}+\sigma_{-}\right)\right)t\right ]
\eeq

In stationary situation, putting the expectation values of the Pauli matrices in Eqn.(\ref{17}), we get the time evolution as

\beq\label{18}
U_s(t)=\exp\left[-i\left(\frac{\Omega}{2}\langle\sigma_z\rangle-\frac{\Delta}{2}\left(\langle\sigma_{+}\rangle+\langle\sigma_{-}\rangle\right)\right)t\right ]
\eeq

Exponent of the time evolution operator

\beq\label{19}
\begin{array}{ll}
\Phi=-it\left[\frac{\Omega}{2}\langle\sigma_z\rangle-\frac{\Delta}{2}(\langle\sigma_{+}\rangle+\langle\sigma_{-}\rangle)\right]\\

=it\left[\frac{\Omega}{2}\frac{(2N(\Omega)+1)}{(2N(\Omega)+1)^2+2|\Lambda|^2/g^2\Omega^2}+\frac{2i\Delta Im(\Lambda)/g\Omega}{(2N(\Omega)+1)^2+2|\Lambda|^2/g^2\Omega^2}\right]\\
~~~~~=\left[i\alpha t-\Gamma t\right]
\end{array}
\eeq
where

\beq\label{20}
\alpha=\frac{\Omega}{2}\frac{(2N(\Omega)+1)}{(2N(\Omega)+1)^2+2|\Lambda|^2/g^2\Omega^2}
\eeq

and

\beq\label{21}
\Gamma=\frac{2\Delta Im(\Lambda)/g\Omega}{(2N(\Omega)+1)^2+2|\Lambda|^2/g^2\Omega^2}
\eeq

Here $\Gamma$ represents the exponential decay of the survival probability for the quantum state. So we can define it to be the decay parameter for the concerning system interacting with the thermal field environment. Here, the decay constant depends on the the coupling parameter $g$, the environmental field variable $\Lambda$ and Planck function $N(\Omega)$, along with the system frequency $\Omega$. Putting the expression of Planck function in (\ref{21}), we get the temperature dependence of the decay parameter $\Gamma$ as

\beq\label{22}
\Gamma=\frac{2\Delta Im(\Lambda)/g\Omega}{\coth^2\left(\frac{\hbar\Omega}{2KT}\right)+2|\Lambda|^2/g^2\Omega^2}
\eeq
From the previous equation (\ref{22}), it follows that

\beq\label{22a}
\Gamma= \frac{1}{\Pi_{th}+\Pi_{q}}
\eeq
where

\beq\label{22b}
\begin{array}{ll}
\Pi_{th}=\frac{g\Omega}{2\Delta Im(\Lambda)}\coth^2\left(\frac{\hbar\Omega}{2KT}\right)\\
\Pi_{q}=\frac{|\Lambda|^2}{g\Omega \Delta Im(\Lambda)}
\end{array}
\eeq

As $T\rightarrow 0$, $\coth\left(\frac{\hbar\Omega}{2KT}\right)\rightarrow 1$. So the decay parameter at zero temperature

\beq\label{23}
\Gamma_0=\frac{2\Delta Im(\Lambda)/g\Omega}{1+2|\Lambda|^2/g^2\Omega^2}
\eeq

\section{Formulation of the weak value of dwell time}

In this section, we will derive the expression of dwell time for our concerning system in the framework of weak measurement. Weak values of a certain operator are the outcome of measurement done on an ensemble of pre and post selected system, where the interaction of the system with the measurement device is sufficiently weak \cite{9a,9b,9c}. In our case, measurement means interaction with the environmental thermal magnetic modes. Due to the interaction with the environmental magnetic modes, the majority of states in the Hilbert space of the concerning system become highly unstable to the interacting environment. So after a short period of time, the system decays into a particular state, which can be decomposed into a mixture of simple pointer states. In our case, the pointer state is the particular state with the characteristic frequency $\Omega$. Because of the reason that the interaction with the magnetic modes is considered to be weak, we are taking the notion of weak measurement to derive the dwell time. For that, we are following the procedure based on one of our recent work \cite{10}, where we have formulated the weak value of dwell time in a dissipative environment on the basis of the procedure of Aharonov et.al \cite{8}.\\
Dwell time is defined as the time interval for the particle residing within the barrier region. An operator can be constructed to determine whether the particle is within the barrier region or not, as

\beq\label{2.1}
\Theta_{(0,L)} = \Theta(x)-\Theta(x-L)
\eeq

where $\Theta$ is heaviside function and $L$ is the width of the magnetic barrier. It gives the values

\beq\label{2.2}
\Theta_{(0,L)}= \left \{  \begin{array}{ll}
                 1 & \mbox{if}~~~ 0<x<L\\
                 0 & \mbox{otherwise}
                 \end{array} \right.
\eeq

The thermal magnetic field is acting within the barrier region $0<x<L$. So the Hamiltonian for the system can be expressed as

\beq\label{2.2a}
H_C= H_T \Theta_{(0,L)}
\eeq

\noindent where $H_T$ is given by (\ref{3}).\\
Now for an observable $A$, if we divide the measurement into many short intervals ($\delta t$), then the weak value of $A$ ($ A\simeq \sum_{j=-\infty}^{\infty} A_j  $), over the ensemble of pre and post selected states $|\psi_i\rangle$ and $|\psi_f\rangle$ respectively, will be given by \cite{8}

\beq\label{2.3}
<A_j>^w=C\Delta t \frac{\langle\psi_f(j\Delta t)|A|\psi_i(j\Delta t)\rangle}{\langle\psi_f(j\Delta t)|\psi_i(j\Delta t)\rangle}
\eeq

where $C$ is an arbitrary constant, which can be set as $C=\frac{1}{\delta t}$. For $\delta t \rightarrow 0$, the summation can be replaced by integration as

\beq\label{2.4}
<\tau>^w= \frac{\int_{-\infty}^{\infty}dt\int_0^L \psi_f^{*}(x,t) \psi_i(x,t)dx}{\int_{-\infty}^{\infty} \psi_f^{*}(x,0) \psi_i(x,0)dx}
\eeq

For a time evolving system, we define the time evolution operator as
\beq\label{2.4b}
U(t-t_i)=e^{-iH_s (t-t_i)}
\eeq

If the initial Hamiltonian is defined as (\ref{1}), the time evolution operator looks like

\beq\label{2.4c}
U(t)= \left (  \begin{array}{ll}
                 e^{i\Omega t/2} & 0\\
                 0 & e^{-i\Omega t/2}
                 \end{array} \right)
\eeq

The initial state, polarized in $x$ direction is given by
\beq\label{2.4d}
|\psi_i\rangle= \frac{1}{\sqrt{2}}\left (\begin{array}{ll}
                         1\\
                         1
                 \end{array} \right)
\eeq

The projection operator onto this particular eigenstate is

\beq\label{2.4e}
P_{+}= \frac{1}{\sqrt{2}}\left (  \begin{array}{ll}
                 1 & 1\\
                 1 & 1
                 \end{array} \right)
\eeq
Let us now consider the case of decay of this particular initial state with the interaction with the bath modes (in this case the thermal magnetic field). Due to this interaction, the initial state is loosing energy to the magnetic bath modes. Let us set the excited states $E_n$ to satisfy the relation
\beq\label{2.4f}
E_n-E_0=n\Delta E,~~~~~-N\leq n\leq N
\eeq
The excited states are chosen to be equispaced and also distributed symmetrically about the excited state of the reference atom, which is taken as $n=0$. The Schr\"{o}dinger equation for the system can be equivalent to the coupled differential equation \cite{15}
\beq\label{2.4g}
\begin{array}{ll}
\dot{a_0}=-i\sum_n H_s a_n e^{-in\Delta Et} \\
\dot{a_n}=-iH_s a_0 e^{in\Delta Et}
\end{array}
\eeq
where $a_n , a_0$ are the amplitudes of the respective states. We should also mention that the coupled differential equations are for the reduced system. According to Davies \cite{15}, solving the coupled set (\ref{2.4g}) by the well known method of Laplace transformation, we get
\beq\label{2.4h}
a_0(t)= e^{-\Gamma (t-t_i)}
\eeq
where $\Gamma$ is the decay parameter. The effect of the environmental interaction is included in this parameter. The time evolution operator for the relevant sub-space of the complete Hilbert space can be represented by a $(2N+1)\times(2N+1)$ dimensional matrix. The components of the matrix are calculated from (\ref{2.4g}) as
\beq\label{2.4i}
U_{00}=e^{-\Gamma t}
\eeq

\beq\label{2.4j}
U_{n0}=iH_s\left[\frac{e^{-\Gamma t+in\Delta Et}-1}{\Gamma -in\Delta E}\right ]
\eeq
in the limit $\Delta E\rightarrow 0$.
For the case of time dependent decaying states \cite{15}, using the relation $U^{\dagger}(t)=U(-t)$, the weak value of a certain operator $A$, can be defined as
\beq\label{2.4a}
A_w= \frac{\langle\psi_f|U^{\dagger}(t-t_f)AU(t-t_i)|\psi_i\rangle}{\langle\psi_f|U^{\dagger}(t-t_f)U(t-t_i)|\psi_i\rangle}
\eeq
If we chose $A$ as the projection operator $P_{+}$ onto the excited state at a certain instant $t$, with the condition that it is pre-selected in the excited state at the instant $t_i$ and after decay, post selected at the instant $t_f$. Let us choose the possible final state as
\beq\label{2.4k}
|\psi_f\rangle=|\psi_k\rangle
\eeq
After some simple calculations \cite{15}, it has been shown that the weak value of the projection operator $P_{+}$ takes the form
\beq\label{2.4l}
P_w=\frac{U_{k0}(t_f-t)U_{00}(t-t_i)}{U_{k0}(t_f-t_i)}
\eeq
Taking the components of the $(2N+1)\times(2N+1)$ dimensional time evolution matrix, as given by (\ref{2.4i}) and (\ref{2.4j}), we get
\beq\label{2.4m}
P_w=e^{-\Gamma(t-t_i)} \left[\frac{1-e^{-\Gamma(t_f-t)+ik\Delta E(t_f-t)}}{1-e^{-\Gamma(t_f-t_i)+ik\Delta E(t_f-t_i)}}\right]
\eeq
For the specific case of $E_k=E_0$, the expression reduces to
\beq\label{2.4A}
P_w=e^{-\Gamma(t-t_i)} \left[\frac{1-e^{-\Gamma(t_f-t)}}{1-e^{-\Gamma(t_f-t_i)}}\right]
\eeq
where $t_i$ and $t_f$ are the time instant for initial and final measurement. $\tau_m(=t_f-t_i)$ can be considered as the measurement time. Since the initial pre-selected excited state is $E_0$, the specific choice of post-selection $E_0$ gives us the weak value of survival probability for the pre-selected state. The time integral of this weak survival probability is realized as the weak value of dwell time \cite{10}. This understanding conforms with the interpretation of dwell time \cite{3,4,5,6}. The barrier region can be understood as a kind of capacitive region, which accumulates and scatters energy (or particle) incident upon it. So the dwell time is nothing but the lifetime of energy storage in the barrier region. Consequently in our case, the projection operator $P_{+}$ can be interpreted as the operator $\Theta_{(0,L)}$. Following this argument, the time integral of the weak value of survival probability can be interpreted as the weak dwell time. If we take into account the superposition of all the excitations of the bath modes, then (\ref{2.4A}) will be somewhat modified \cite{10,15}. But that is not necessary in this particular case, because here the system is in resonance with one particular mode (with frequency $\Omega$) and the effect of all others is negligible. Following this argument, the weak dwell time is defined as
\beq\label{2.5}
\tau_D^w=\int_{t_i}^{t_f} e^{-\Gamma(t-t_i)} \left[\frac{1-e^{-\Gamma(t_f-t)}}{1-e^{-\Gamma(t_f-t_i)}}\right]dt
\eeq
Since the pre and post selection are done at the time instant $t_i$ and $t_f$, we have taken the lower and upper limit as those two time instants respectively. Calculating further from Eqn. (\ref{2.5}) we get
\beq\label{2.6}
\tau_D^w=\frac{1}{\Gamma}\left[1-\frac{\Gamma\tau_m}{e^{\Gamma\tau_m}-1}\right]
\eeq
Now, the measurement time ($\tau_m$) is defined as the time interval between two successive interactions of the system with the magnetic field. We can assume it to be frequent enough for holding the inequality $\tau_m\ll 1/\Gamma$. So under this assumption, if we take the exponential term up to 2nd order, the dwell time reduces to be
\beq\label{2.7}
\tau_D^w=\frac{1}{2/\tau_m+\Gamma}
\eeq
The amplitude of the magnetic field varies with it's characteristic frequencies (in this case we have only considered the resonant frequency $\Omega$, since the effects of other frequencies are negligible). After the time period which equals to the inverse of this frequency, the amplitude of the magnetic field becomes maximum, which can also be taken as the instant of maximum interaction between the system and the field. So it is perfectly plausible to take the time interval between two successive measurements ($\tau_m$) to be equal to the inverse of the resonant frequency ($\Omega$) of the system and the field. Then the dwell time becomes
\beq\label{2.7}
\tau_D^w=\frac{1}{2\Omega+\Gamma}
\eeq

Now at this point of our discussion, one can also ask that why the Dwell time is not dependent on the spacial parameter like barrier width. To answer this question, we must firstly point out that here we are dealing with the time evolution of the system. So the parameters on which the time scale should depend are of the dimension of corresponding conjugate quantity, ie. frequency (or energy). But more importantly, in this work we do not consider the process of tunneling as a process of the wave packet traversing through the barrier region. Actually the barrier acts like a sort of capacitive region which accumulates energy which falls upon it in the form of incident wave packet and after a time delay it scatters the energy from both sides \cite{3,4,5,6}. The part of energy scattered on the other side of the barrier is the tunneling part. The time delay between the the accumulation and scattering is understood as the Dwell time.\\
To observe the temperature dependence of the dwell time, we put the expression of $\Gamma$ form Eqn. (\ref{22a}) in Eqn.(\ref{2.7}) to get
\beq\label{2.8}
\tau_D^w=\frac{\Pi_{th}+\Pi_{q}}{1+2\Omega(\Pi_{th}+\Pi_{q})}
\eeq
where $\Pi_{th}$ corresponds to the effect of temperature in the expression of dwell time. The variation of $\tau_D^w$ with increasing temperature is shown in Fig.1.

\begin{figure}[htb]
{\centerline{\includegraphics[width=7cm, height=5cm] {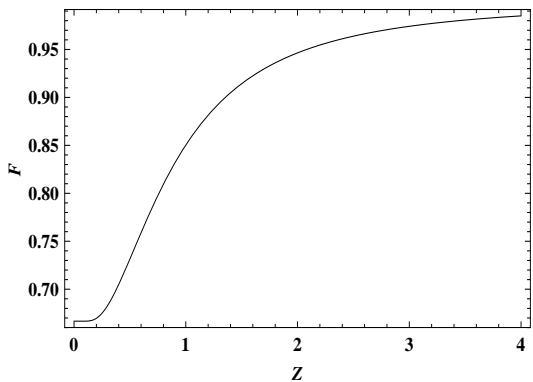}}}
\caption{$F$ vs. $z$. where $z=KT/\hbar\Omega$ and $F=2\Omega\tau_D^w$. Since $\Omega$ has a fixed value, the graph basically shows the variation of $\tau_D^w$ with $z$}
\label{figVr}
\end{figure}
From the figure, we can clearly see that dwell time initially increases with the increase of temperature and finally reaches a saturation value at high temperature. The minimum value of the dwell time is at zero temperature, which is given by
\beq\label{2.9}
\tau_{D0}^w=\frac{\Pi_{q}(1+g^2\Omega^2/|\Lambda|^2)}{1+2\Omega\Pi_q(1+g^2\Omega^2/|\Lambda|^2)}
\eeq
\section{Conclusion}

In this work, we have discussed the time evolution of a two level system coupled to a thermal magnetic noise field environment and evaluated the dwell time for a two level particle tunneling through the barrier. As we have stated earlier, the practical example of the situation is of TEM, where the noisy thermal magnetic field acts as the cause of decoherence. Our analysis shows that, dwell time increases with the increase of temperature and finally reaches a saturation value at high temperature. The increment of dwell time with rising temperature can be interpreted as the loss of memory of original tunneling direction caused by efficient energy exchange between the system and the environmental modes. With the rise of temperature, as the process becomes kinematically more and more random, this effect increases and so does the dwell time. We have also shown that at sufficiently high temperature the dwell time saturates, because at such high temperature, the process does not remain quantum mechanical at all. This is simply the case of thermal hopping over the barrier. As the temperature decreases, the quantum effect becomes more and more important. It is also worth mentioning that, from Eqn. (\ref{22}) we can see that with the rise of temperature the decay parameter decreases. This is because of the fact that for increasing environmental temperature, the possibility for the particle to absorb sufficient energy to remain in it's initial state also increases. Thus we infer that the rise of environmental temperature can preserve the quantum state to a certain extent. In this context, it is worth mentioning that in a recent work \cite{16}, it has been shown that temperature can strengthen the effective coupling between quantum states and hence can be responsible for occurrence of some Zeno phenomena. We conclude with the note that it is our aim to extend our formalism to the aspect of Zeno dynamics in subsequent publications.

\section{Acknowledgement}

The author thanks Prof. Sisir Roy of Indian Statistical Institute for helpful discussions.

\end{document}